\documentclass[lettersize]{IEEEtran}
\usepackage{amsmath,amsfonts}
\usepackage{algorithmic}
\usepackage{algorithm}
\usepackage{array}
\usepackage[caption=false,font=normalsize,labelfont=sf,textfont=sf]{subfig}
\usepackage{textcomp}
\usepackage{stfloats}
\usepackage{url}
\usepackage{verbatim}
\usepackage{graphicx}
\usepackage{cite}
\usepackage{afterpage}
\usepackage{float}
\usepackage{xspace}

\newcommand{\X}{$\times$\xspace}

\begin{document}

\title{\textbf{Flex-PE}: Flexible and SIMD Multi-Precision Processing Element for AI Workloads}

\author{

Mukul Lokhande,
Gopal Raut,~\IEEEmembership{Member, IEEE,}
Santosh Kumar Vishvakarma,~\IEEEmembership{Senior Member, IEEE,} 
        
\thanks{
This work is supported under the Special Manpower Development Program for Chip to Startup (SMDP-C2S), Ministry of Electronics and Information Technology (MeitY), Govt. Of India.
\\
The authors are associated with the NSDCS Research Group, Department of Electrical Engineering, IIT Indore, Simrol-453552, India.
\\
\textbf{Corresponding author}: Santosh Kumar Vishvakarma.\\
\textbf{E-mail:} skvishvakarma@iiti.ac.in.
}
\thanks{Manuscript received ; revised .}}

\markboth{PREPRINT: IEEE Transactions on Very Large Scale Integration (VLSI) Systems, ~Vol.~XX, No.~X, XXX~202X}%
{Lokhande \MakeLowercase{\textit{et al.}}: Flexible and SIMD Multi-Precision Processing Elements for AI Workloads}


\maketitle

\begin{abstract}
The rapid adaptation of data-driven AI models, such as deep learning inference, training, Vision Transformers (ViTs), and other HPC applications, drives a strong need for run-time precision configurable different non-linear activation functions (AF) hardware support. Existing solutions support diverse precision or run-time AF reconfigurability but fail to address both simultaneously. 
This work proposes a flexible and SIMD multi-precision processing element (Flex-PE), which supports diverse run-time configurable AFs, including sigmoid, tanh, ReLU and softmax, in addition to MAC operation. The proposed design achieves an improved throughput of up to 16\X FxP4, 8\X FxP8, 4\X FxP16 and 1\X FxP32 in pipeline mode with 100\% time-multiplexed hardware. 
This work proposes an area-efficient multi-precision iterative mode in the SIMD systolic arrays for edge-AI use cases. The design delivers superior performance with up to 62× and 371× reductions in DMA reads for input feature maps and weight filters in VGG-16, with an energy efficiency of 8.42 GOPS / W within the accuracy loss of 2\%.
The proposed architecture supports emerging 4-bit computations for DL inference while enhancing throughput in FxP8/16 modes for transformers and other HPC applications. The proposed approach enables a way for future energy-efficient AI accelerators in both edge and cloud environments. 

\end{abstract}


\section{Introduction}

\begin{figure}
\centering
\subfloat[\label{general-DNN-arch}]{\includegraphics[width=\columnwidth,height=37.5mm]{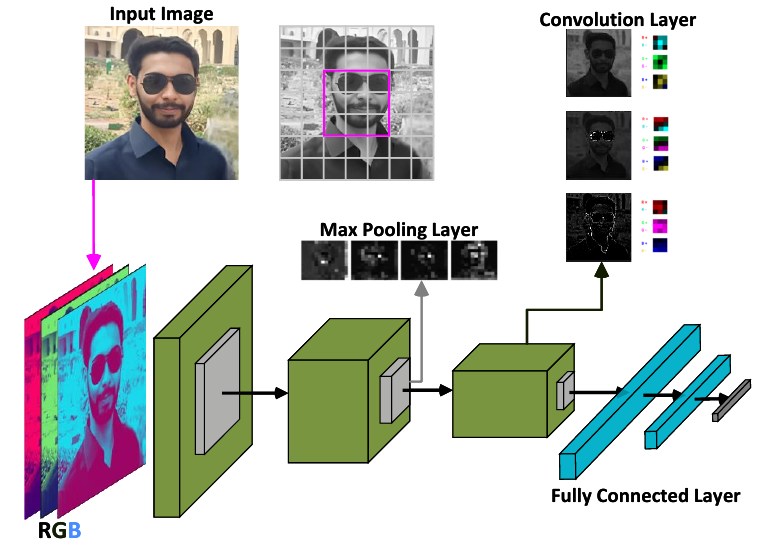}%
\vspace{-5 mm}
}
\hfill
\subfloat[\label{Edge-AI-SoC}]{\includegraphics[width=\columnwidth,height=40mm]{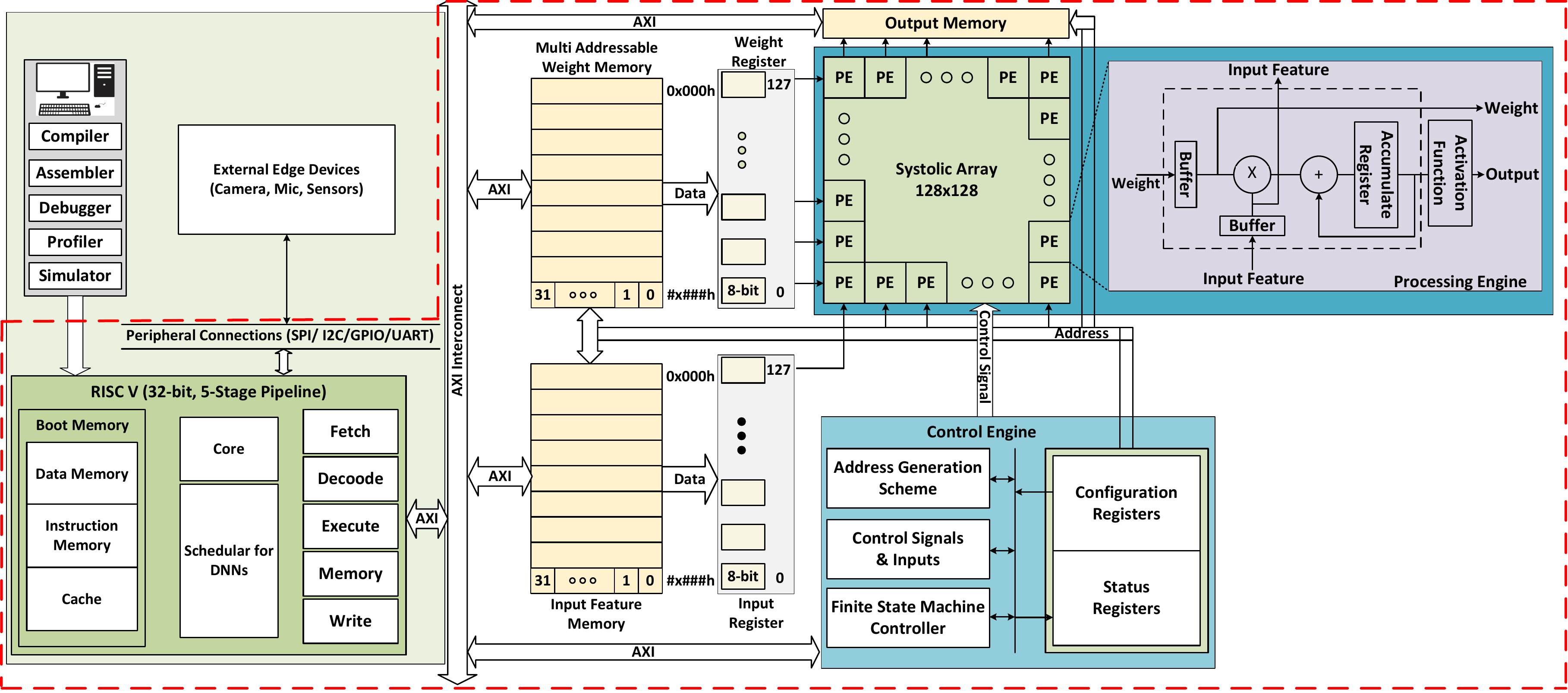}}%
\caption{(a) Typical Deep Neural Network (DNN) model showcasing various layers including Conv, Pooling, and FC.
(b) AI SoC featuring a RISC-V-enabled Systolic Array with detailed PE architecture.}
\vspace{-5 mm}
\end{figure}

\begin{figure}[!t]
    \centering
    \includegraphics[width=\linewidth]{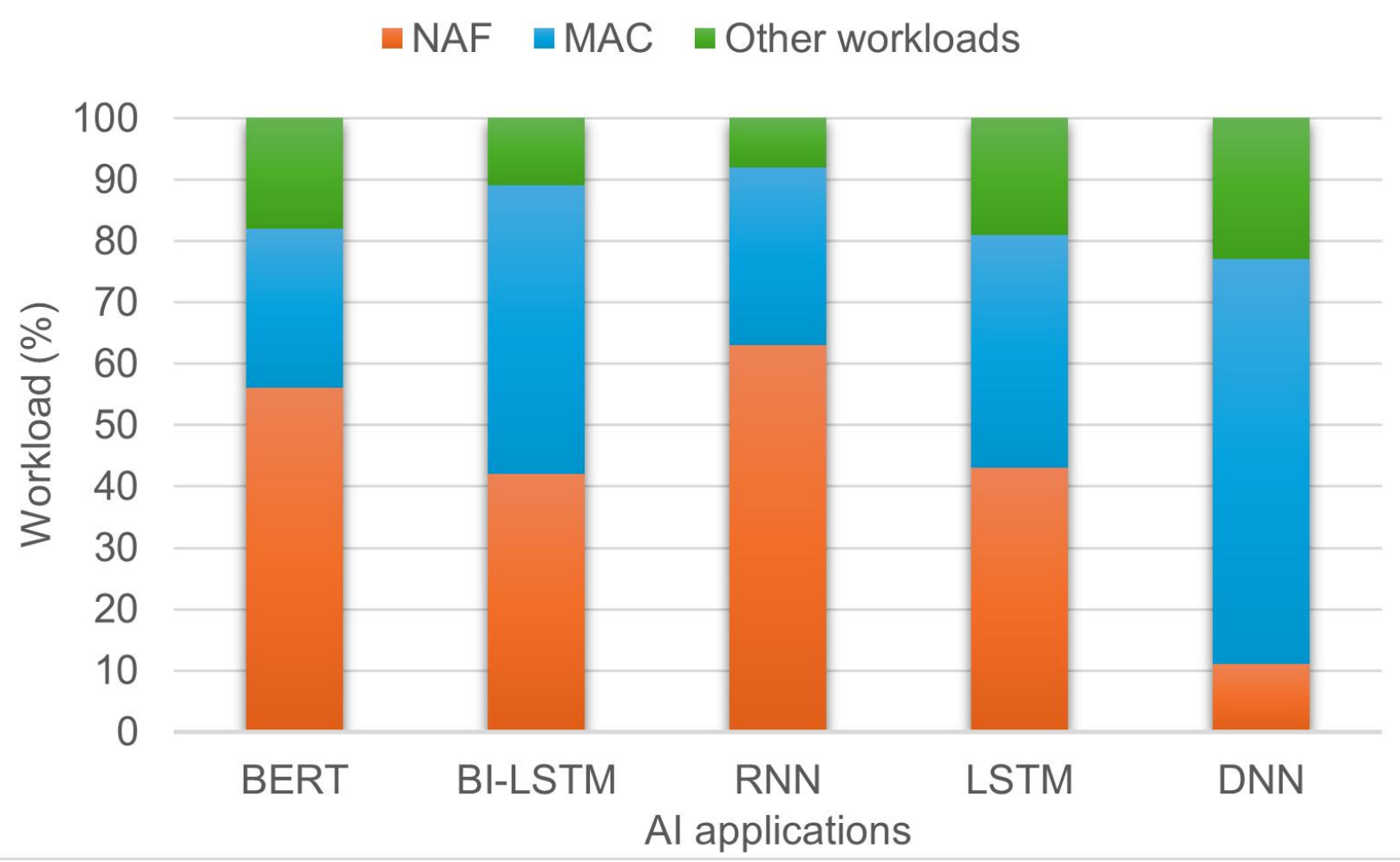}
        \caption{Workload analysis\cite{TAI24-CORDIC-RNN} emphasizing on the growing demand for performance-enhanced non-linear activation functions.}
    \label{AI-workload}
    \vspace{-2mm}
\end{figure}

\IEEEPARstart{A}{rtifical} Intelligence (AI) has grown far ahead of human capabilities and is a major data-driven decision maker.
However, the demand for computing resources has grown rapidly from deep neural networks (DNNs) to the recent large-language models (LLMs).
This highlights the need for hardware accelerators.
We must not only focus on increasing the peak compute power of hardware but also keep in mind memory usage and intra/inter-chip communication bottlenecks while serving AI models~\cite{AI-Memorywall24}. 
In the past 20 years, TOPS performance has been enhanced by 60,000\X, while DRAM bandwidth and interconnect bandwidth have been improved by only 30\X and 100\X respectively, creating a disparity known as the memory wall problem~\cite{AI-Memorywall24}.
The memory wall problem is characterized by limitations in memory capacity, data transfer bandwidth, and access latency, which majorly affect overall system performance. 
The bottlenecks have not been only for edge inference evaluation on real-time data but also for cloud training with collected data.
Thus, extensive parallel processing with quantised precision has become the need of the hour, considering the rapid growth in AI use cases to exploit throughput gains.
AI applications, from real-time edge inference to large-scale cloud training, have forced the necessity for hardware accelerators that can effectively trade between performance, flexibility, and efficiency. The rapidly varying diversity activation functions (AFs) and precision support introduce an additional layer of complexity in the design for such accelerators. The breakdown of the total execution of different AI workloads, including DNN (LeNet), RNNs (LSTM, BI-LSTM, and GRU) and the transformer (BERT), Fig. \ref{AI-workload} highlights the significant need for AF optimization. 

Edge-AI devices require rapid event-triggered response, superior operational efficiency, and dense computational throughput. In contrast, high-performance computing (HPC) devices focus on providing enhanced large-scale performance within available memory bandwidth and interconnect at the cloud node. The general DNN hardware architecture is shown in Fig. 1(a), and RISC-V enabled AI SoC is depicted in Fig. 1(b) with Systolic array architecture. 
Previous works have explored quantization and Single Instruction Multiple Data (SIMD) vectorization techniques to provide efficient trade-offs between arithmetic intensity and data feed rates~\cite{SIMD-Vector-ProcArch}. Precision-reduction techniques like quantization enable faster inference with minor accuracy compromise, while high-precision training burdens computational and memory resources, forcing bandwidth bottlenecks.
Prior works have demonstrated consumer-satisfactory DNN and Transformer inference within 4 or 8-bit dynamic fixed-point (FxP) precision. 
However, higher precision is necessary for handling error accumulation, precise gradient calculations, critical data dependencies, attention fidelity and convertible backpropagation in DNN training and HPC applications such as RNNs, LSTMs, and Transformers to achieve satisfactory performance~\cite{Embedded-DL-accl, FxP-Transformers, Systolic-Trans-Fxp-TAI24}. 
Pre-processing blocks can easily handle the conversion between various intermediate precision or integer to fixed-point conversions~\cite{Reconfigurable-PE-TCASII24} without contributing any overhead.

\begin{table*}[t]
\caption{State-of-the-art methodologies and comparison of corresponding features in AF units for AI applications.}
\label{SOTA-comp}
\resizebox{\textwidth}{!}{%
\begin{tabular}{|c|cccccccc|c|}
\hline
\textbf{Comparison} & \textbf{AF supported} & \textbf{Configurability} & \textbf{SIMD supported} & \textbf{Supported Precision} & \textbf{Datatypes} & \textbf{Throughput} & \textbf{Hardware architecture} & \textbf{\begin{tabular}[c]{@{}c@{}}Design Overhead \\ (Area/Delay)\end{tabular}} & \textbf{Applications} \\ \hline
\textbf{TAI-24~\cite{TAI24-CORDIC-RNN}} & Sigmoid, Tanh & Yes & No & 24 & Fixed & 1/1/1/1/1/1 & Pipelined & Area & RNN/LSTM \\ \hline
\textbf{TCSVT'24~\cite{TCSVT24_SoftAct}} & Softmax & No & No & 16/32 & Fixed & 1/1 & pipeline & Area & Transformers \\ \hline
\textbf{ISQED'24~\cite{MRao-ISQED24}} & Sigmoid, Tanh, Softmax & No & No & 12/16/32 & Bf/Tf/Float/Posit & 1/1/1 & Iterative/Parallel/Pipelined & Delay/Area & edge inference \\ \hline
\textbf{ISQED'24~\cite{GR-ISQED24}} & Sigmoid, Tanh & Yes & Yes & 8/16 & Fixed & 2/1 & Pipelined & Area & DNN inference \\ \hline
\textbf{TCAS-I'23~\cite{TCASI23-Softmax}} & Softmax & No & No & 32 & Fixed & 1 & Pipelined & Area & DNN Training \\ \hline
\textbf{TCAS-II'23~\cite{TCASII-23_ReAFM}} & Sigmoid, Tanh, Swish, ReLU & Yes & No & 12 & Fixed & 1 & Taylor-Series + NRDiv & NA & DNN inference \\ \hline
\textbf{IEEE Access'23~\cite{Softmax-Sumi}} & Softmax & No & No & 8/16/32 & Fixed & 1/1/1 & Pipelined & Area & DNN inference \\ \hline
\textbf{TC'23~\cite{TC23-CORDIC-LSTM}} & Sigmoid, Tanh & Yes & No & 16 & Fixed & 1 & Pipelined & Area & LSTM \\ \hline
\textbf{ISVLSI'20~\cite{GR-configAF-ISVLSI20}} & Sigmoid, Tanh & Yes & No & 8/12/16 & Fixed & 1/1/1 & Iterative & Delay & ANN \\ \hline
\textbf{TCAD'19~\cite{TCAD19-AF}} & Tanh, SELU & No & No & 7/8 & Fixed & 1/1 & LUT-combo & NA & ANN \\ \hline
\textbf{Proposed} & Sigmoid, Tanh, Softmax, ReLU & Yes & Yes & 4/8/16/32, 12/24* & Fixed & 16/8/4/1 & Flexible & Runtime tradeoff & \begin{tabular}[c]{@{}c@{}}DNN, RNN/LSTM, Transformers \\ Training and Inference\end{tabular} \\ \hline
\end{tabular}}
\vspace{1mm}
\par \textbf{Note}\textsuperscript{$\ast$}: The work also supports 2$\times$FxP12, 4$\times$FxP4 or 1$\times$FxP24 heterogeneous precision operations.
\vspace{-5mm}
\end{table*}

AI compute hardware primarily consists of general matrix-matrix multiplication (GEMM), matrix-vector multiplication (MVM), and multiply-and-accumulate (MAC) operations to execute diverse DL workloads-such as convolution, recurrent and fully connected (FC) layers and Transformer workloads, involving encoder and decoder layers. Existing accelerators are effective for specific tasks but lack the flexibility to support dynamically varying precisions and activation function requirements, limiting the execution of edge-AI and HPC workloads simultaneously.
Prior works~\cite{Intel-MxCore} highlight that significant matrix processing cores are necessary for highly parallelizable workloads, but the performance is bottlenecked by challenges associated with thread scheduling, data moment, resource utilisation and the energy efficiency of the varying proportion of scalar, non-MAC operations. The CORDIC-based methodology has been widely used in the previous methodologies\cite{GR-ACM_TRETS23, GR-configAF-ISVLSI20}. AI algorithm-specific operand dependency-aware tiling overcomes these limitations by enhancing compute utilization, system performance, and power efficiency\cite{slimmerCNN}. 
However, the efficient SIMD-enabled hardware is crucial to fully leverage these optimizations and execute these operations in parallel. 
The approach empowers low latency at the edge node while providing flexibility at the cloud node for workload mapping, optimizing compute density and reducing utilisation of memory bandwidth and interconnect usage.

A key drawback in state-of-the-art (SoTA) works\cite{GR-configAF-ISVLSI20, TCSVT24_SoftAct, TAI24-CORDIC-RNN, MRao-ISQED24, GR-ISQED24, TCASI23-Softmax, TCASII-23_ReAFM, Softmax-Sumi, TC23-CORDIC-LSTM, TCAD19-AF}, which can be observed from Table \ref{SOTA-comp}, is the absence of SIMD-configurable multi-precision activation function (AF) hardware capable of supporting the needs of diverse workloads such as DNN, RNN/LSTM, and Transformer models. 
This work proposes a versatile area-efficient Flex-PE that supports run-time flexibility across diverse activation functions such as Tanh, ReLU, Softmax, and Sigmoid while supporting multi-precision (4/8/16/32) operations at improved throughput. 
The run-time precision switching allows the system to adjust bit-width dynamically and contribute to resource savings at edge inference while minimizing memory footprint and power consumption in the cloud while preserving accuracy. 
The configurable activation function supports different non-linear transformations for diverse workload needs with the same hardware, embracing resource efficiency for various applications.
Combined, these advancements enable the development of general AI hardware accelerators with efficient resource utilisation and enhanced performance for more adaptability to the rising demands of complex workloads.

The major contributions of this work are:







\begin{itemize}

    \item \textbf{Flexible multi-precision Reconfigurable activation function (AF):}
    \begin{itemize}
        \item The proposed Flex-PE supports diverse activation functions (Tanh, ReLU, Softmax, Sigmoid) with multi-precision versatility (FxP4/8/16/32).
        \item The proposed PE achieves enhanced throughput (16/8/4/1 for FxP4/8/16/32) with time-multiplexing and almost 100\% hardware utilization.
        \item The proposed work explores the configurable FxP4 PE for edge-AI applications beyond ReLU.  
    \end{itemize}

\newpage
    \item \textbf{Flexible and SIMD multi-precision CORDIC-based processing element (Flex-PE):}
    \begin{itemize}
        \item The proposed Flex-PE use the CORDIC methodology for MAC and diverse AF operations to deliver a reconfigurable area-efficient AI core.
        \item The proposed Flex-PE features an iterative mode for resource-constrained environments and a pipelined mode for HPC applications, achieving a trade-off between latency and application performance. 
        \item The proposed Flex-PE adapts the diverse needs of AI workloads from DL inference \& training, RNN/LSTM, and Transformers to HPC workloads.
    \end{itemize}

    \item \textbf{Analysis for performance-enhanced SIMD systolic array-based on proposed Flex-PE:}
    \begin{itemize}
        \item The performance-enhanced SIMD systolic array has been evaluated for edge applications with runtime adaptability between precision levels (FxP4/8/16/32) and FPGA hardware resources to achieve a throughput of 8.42 GOPS/W within 2\% accuracy loss.  
        \item The proposed hardware delivers superior performance for VGG-16, with optimal precision within available memory bandwidth, with up to 62$\times$ and 371$\times$ reductions in DMA reads for inputs feature and weight filters through the SIMD data flow scheduler.

    \end{itemize}
\end{itemize}

The rest of this paper is structured as follows:
Section II analyses AF design methodology and current SoTA works. 
Section III outlines the architecture of the proposed SIMD PE. 
Section IV describes the implementation methodology and performance evaluation results, and Section V concludes.

\section{The novel SIMD Configurable Activation Function and Theoretical analysis}

\subsection{Previous Design methodologies}

AI workloads consist of different layers, such as convolutional, recurrent, fully connected, pooling, LSTM, normalization, and activation layers, including ReLU, tanh, and sigmoid. The workload analysis has been detailed in Fig. \ref{AI-workload}
The core of these workloads is processing elements (PEs/Neurons)~\cite{RASHT, Reconfigurable-PE-TCASII24}, which are the critical drivers of computational efficiency. 
A reconfigurable multi-precision PE is recommended to enhance the system performance, typically providing multi-precision computations for the operations involved.
 
The run-time precision switching allows the system to adjust bit-width dynamically and contribute to resource savings at edge inference while minimizing memory
footprint and power consumption in the cloud while preserving accuracy\cite{LUT-AF}.
The configurable activation function supports different non-linear transformations for diverse workload needs with the same hardware, embracing resource efficiency for various applications.
The detailed comparison of SoTA AF design methodologies with precision supported and reconfigurability, primarily emphasising AFs such as ReLU, Exp, Tanh, Sigmoid and Softmax, is discussed in 
Table \ref{SOTA-comp}. The mathematical equations are defined as in Eq. [\ref{eq_AF}]:

\begin{equation} \label{eq_AF}
    \begin{array}{ll}
        \text{Exponential:} & e^x = \sinh(x) + \cosh(x)\\[5pt]
        \text{Sigmoid:} & \sigma(x) = \frac{e^{x}}{1 + e^{x}} \\[5pt]
        \text{Tanh:} & \tanh(x) = \sinh(x) / \cosh(x)\\[5pt]
        \text{ReLU :} & \text{ReLU}(x) = \max(0, x) \\[5pt]
        \text{Softmax:} & \text{Softmax}(x_i) = \frac{e^{x_i}}{\sum_{j} e^{x_j}}.
    \end{array}
\end{equation}

Combined, these advancements not only enhance the adaptability of the hardware, but also optimize resource efficiency, making it highly suitable for complex AI applications.
To address wide dynamic ranges and potential quantization errors, prior works have proposed solutions for multi-precision neural processing units~\cite{Intel-MxCore, Systolic-PE_TCASI24}, but these works do not discuss the multi-precision AF computations. 
This marks the primary motivation for our work.

\begin{table}[!t]
    \caption{CORDIC Hyperbolic Rotational Mode ($\cosh$ \& $\sinh$)}
    \label{hyp-cordic}
    \centering
    \resizebox{\linewidth}{!}{%
    \begin{tabular}{cccccr}
    \hline
    \textbf{i (clk)} & \textbf{Ei = tanh\textsuperscript{-1}(2\textsuperscript{-i})} & \textbf{Xi+1 $\rightarrow$ $\cosh$(0.5)} & \textbf{Yi+1 $\rightarrow$ sinh(0.5)} & \textbf{Zi+1 $\rightarrow$ 0} & \textbf{di} \\ \hline
     &  & \textbf{Initial} Xi = 1/k' & \textbf{Initial} Yi = 0 & Zinput = 0.5 &  \\ \hline
    \textbf{1} & 0.5493 & 1.2075 & \textbf{0.6037} & -0.0493 & 1 \\ \hline
    \textbf{2} & 0.2554 & 1.0566 & 0.3019 & 0.2061 & -1 \\ \hline
    \textbf{3} & 0.1257 & 1.0943 & 0.4339 & 0.0804 & 1 \\ \hline
    \textbf{4} & \textbf{0.0626} & \textbf{1.1214} & \textbf{0.5023} & \textbf{0.0179} & \textbf{1} \\ \hline
    \textbf{5} & 0.0313 & 1.1371 & 0.5374 & -0.0134 & 1 \\ \hline
    \textbf{6} & 0.0156 & 1.1287 & 0.5196 & 0.0022 & -1 \\ \hline
    \textbf{7} & 0.0068 & 1.1328 & 0.5284 & -0.0046 & 1 \\ \hline
    \textbf{8} & 0.0039 & 1.1307 & 0.5240 & -0.0007 & -1 \\ \hline
    \textbf{9} & 0.0020 & \textbf{1.1297} & \textbf{0.5218} & \textbf{0.0013} & -1 \\ \hline
    \end{tabular}}
    \vspace{-3mm}
\end{table}

Prior works~\cite{GR-ACM_TRETS23, GR-Neuro, GR-configAF-ISVLSI20, Intel-taylor_Patent22, NVIDIA-Softmax_Patent22, Intel-ApproxLUT_Patent22_1, Intel-ApproxLUT_Patent22_2} have explored various design approaches for hardware acceleration of non-linear activation functions.
The major methodologies focus on LUT-based approaches that store values or parameters, LUT-based piecewise linear (PWL) approximation, Stochastic computation (SC) techniques, Taylor series approximation\cite{TCASII-23_ReAFM}, and CORDIC-based shift-add.
PWL approximation~\cite{Intel-taylor_Patent22, NVIDIA-Softmax_Patent22, GR-Neuro}, and LUT-based methods~\cite{Intel-ApproxLUT_Patent22_1, Intel-ApproxLUT_Patent22_2} are proven insignificant at higher precision and impact performance due to interpolation and granularity issues.
The bits representing each value, as well as BRAM ports/search latency, limit the multi-precision approach even in FPGA implementations~\cite{GR-Neuro}.
Step artifacts and Quantization errors are introduced in SC and Taylor series approximation approaches~\cite{Intel-taylor_Patent22, RECONFIG-MP-QuantAware-NAF}, and the performance is severely sensitive to segment selection and convergence terms, also proven to be less resource-efficient.
CORDIC methodology is proven to be area-efficient for designing various activation functions, such as Sigmoid~\cite{TAI24-CORDIC-RNN}, Tanh~\cite{TCAD19-AF}, and Softmax~\cite{TCSVT24_SoftAct}.
This method utilises a simple Shift-Add methodology to implement the activation function in an iterative/pipelined fashion and offers an opportunity for simpler high-performance SIMD multi-precision AF hardware design.
Table \ref{SOTA-comp} provides a detailed comparison of SoTA AF design methodologies with precision supported and reconfigurability, primarily emphasizing AFs such as ReLU, Tanh, Sigmoid, and Softmax. Thus, we proceed with the above-mentioned AFs in our approach.

\begin{table}[!t]
    \caption{CORDIC Linear Vectoring Mode (Division)}
    \label{cordic-linear}
    \centering
    \resizebox{\linewidth}{!}{%
    \begin{tabular}{cccccr}
    \hline
    \textbf{i (clk)} & \textbf{Ei=2\textsuperscript{-i}} & \textbf{Xi+1 $\rightarrow$ Xi} & \textbf{Yi+1 $\rightarrow$ 0} & \textbf{Zi+1 $\rightarrow$ Y/X} & \textbf{di} \\ \hline
    \textbf{1}  & 0.5      & 2.51 & -0.734    & 0.5      &-1\\ 
    \textbf{2}  & 0.25     & 2.51 & -0.1065   & 0.25     &1\\ 
    \textbf{3}  & 0.125    & 2.51 & 0.20725   & 0.125    &1\\ 
    \textbf{4}  & 0.0625   & 2.51 & 0.050375  & 0.1875   &-1\\ 
    \textbf{5}  & \textbf{0.03125}  & \textbf{2.51} & \textbf{-0.02806}  & \textbf{0.21875}  &\textbf{-1}\\ 
    \textbf{6}  & 0.015625 & 2.51 & 0.011156  & 0.203125 &1\\ 
    \textbf{7}  & 0.007812 & 2.51 & -0.00845  & 0.210937 &-1\\ 
    \textbf{8}  & 0.003906 & 2.51 & 0.001351  & 0.207031 &1\\ 
    \textbf{9}  & 0.001953 & 2.51 & -0.00355  & 0.208984 &-1\\ \hline
    \end{tabular}}
    \vspace{-5mm}
\end{table}

\subsection{CORDIC computation methodology}

The unified CORDIC algorithm performs circular, linear, and hyperbolic operations with the same hardware in both rotational and vector modes~\cite{Unified-CORDIC, RECON}.
All fundamental mathematical operations shown in Table 1 of article~\cite{Softmax-Sumi} can be performed with the CORDIC approach where planar coordinates are transformed into rotational ones and simple logic blocks, such as Add/Sub, MUX, LBS, and memory elements.
CORDIC performs pseudo-rotation of vectors with two modes, Vectoring and Rotational, further operating with three planar coordinates – Circular, Linear, and Hyperbolic to enumerate arithmetic, trigonometric and complex mathematical functions.
The CORDIC algorithm is a scaled rotation of $X$, $Y$ and $Z$ variables, where $X$ and $Y$ are coordinates of pseudorotation, and $Z$ keeps track of the angle at which the pseudo-vector rotates.

For hardware implementation, the CORDIC trigonometric equations simplify pseudo-rotation to converge into the linear form, as demonstrated in Eq. [\ref{CORDIC-algo}]:
\begin{equation}
    \begin{aligned}
    X_{i+1} &= X_i - m \cdot d_i \cdot Y_i \cdot 2^{-i} \\
    Y_{i+1} &= Y_i + d_i \cdot X_i \cdot 2^{-i} \\
    Z_{i+1} &= Z_i - d_i \cdot E_i
    \end{aligned}
    \label{CORDIC-algo}
\end{equation}

These variables and scaling factor $K$ ($Kh=0.8281$ and $Kc= 1.6467$) are initiated based on the operation mode.
At the $i^{\text{th}}$ iteration, the variables values converge to $X_i$, $Y_i$, and $Z_i$. 
The memory element \( E_i \) is \( 2^{-i} \), \( \tan^{-1}(2^{-i}) \), and \( \tanh^{-1}(2^{-i}) \), and mode \( m \in \{0, 1, -1\} \) for Linear, Circular, and Hyperbolic coordinates, respectively, at each iteration \( i \in \{1, 2, 3, \dots, n\} \).
The rotation direction for the $i$\textsuperscript{th} iteration, \( d_i \in \{1, -1\} \) is determined by  \( \text{sign}(Z_i)\) and  \( -(\text{sign}(X_i)  \oplus  \text{sign}(Y_i)) \) for rotational mode and for vectoring mode operations.
For the proposed methodology, we used Hyperbolic rotational (HR) mode for the evaluation of exponential function, Linear Vector (LV) mode for division mode in AF calculation, and Linear Rotational (LR) mode as MAC to utilize in the same fashion as detailed in ~\cite{RECON, GR-configAF-ISVLSI20, Softmax-Sumi}.

\subsection{CORDIC Hyperbolic mode}\label{sec:cordic-hyperbolic}

For the calculations of hyperbolic trigonometric function $\sinh$ and $\cosh$ required in the evaluations of the exponential function, Sigmoid, Tanh, and Softmax, the CORDIC algorithm is used in HR mode, as shown in Eq. [\ref{CORDIC-hyp}]:
\begin{equation}
    \begin{aligned}
    X_{i+1} &= X_i + d_i \cdot Y_i \cdot 2^{-i} \\
    Y_{i+1} &= Y_i + d_i \cdot X_i \cdot 2^{-i} \\
    Z_{i+1} &= Z_i - d_i \cdot \tanh^{-1}(2^{-i}).
    \end{aligned}
    \label{CORDIC-hyp}
\end{equation}

The algorithm is initiated with scaled-elimination methodology, the initial variables $m$ = --1, $X_0$ = $1/Kh$ = 1.2074, $Y_0$ = 0 and input fed at $Z_0$, with rotations scaling factors as \(\tanh^{-1}(2^{-i})\).
Thus, outputs $X_n$, $Y_n$ and $Z_n$ will converge to $\cosh(Z_i)$, $\sinh(Z_i)$, and zero, respectively, and necessary functions $\sinh$, $\cosh$ are available for further computation.
The detailed mathematical computations for $n$ iterations are shown in Table \ref{CORDIC-hyp}.
The mode of operation in the CORDIC hyperbolic block is very similar as described in~\cite{Softmax-Sumi, GR-configAF-ISVLSI20}.

\subsection{CORDIC linear mode as Division and RECON-MAC}

For the division operation, the CORDIC algorithm is used in LV mode, as showcased in Eq. [\ref{CORDIC-div}]:
\begin{equation}
    \begin{aligned}
    X_{i+1} &= X_i \\
    Y_{i+1} &= Y_i + d_i \cdot X_i \cdot 2^{-i} \\
    Z_{i+1} &= Z_i - d_i \cdot 2^{-i}.
    \end{aligned}
    \label{CORDIC-div}
\end{equation}

The algorithm is initiated with scaled-elimination methodology, the initial variables $m$= 0, $X_0$, $Y_0$ and $Z_0$ as divider (Num), dividend (Denom) and Zero respectively, with memory element as \(2^{-i}\).
Thus, output $Z_n$ shall converge to the quotient, and in our case, Tanh or Softmax, depending on the AF selection mode.
The detailed mathematical computations for $n$ iterations are shown in Table \ref{cordic-linear}.
The linear mode of operation can also be reused as reconfigurable MAC mode as demonstrated in~\cite{RECON}, enabling hardware-efficient reconfigurable PE supporting FxP 4/8/16/32 precision.
The CORDIC computations are limited by a specific convergence range, such as for HR mode \([-1.1182, 1.1182]\), LV mode \([-1, 1]\), and LR mode \([-7.968, 7.968]\).
Thus, to ensure that the input converges, the computation has to be normalised for the range of [\-1,1] and MaxNorm of 5.5, as briefly described in~\cite{GR-ACM_TRETS23}.

\begin{figure}[!t]
    \centering
    \includegraphics[width=\linewidth]{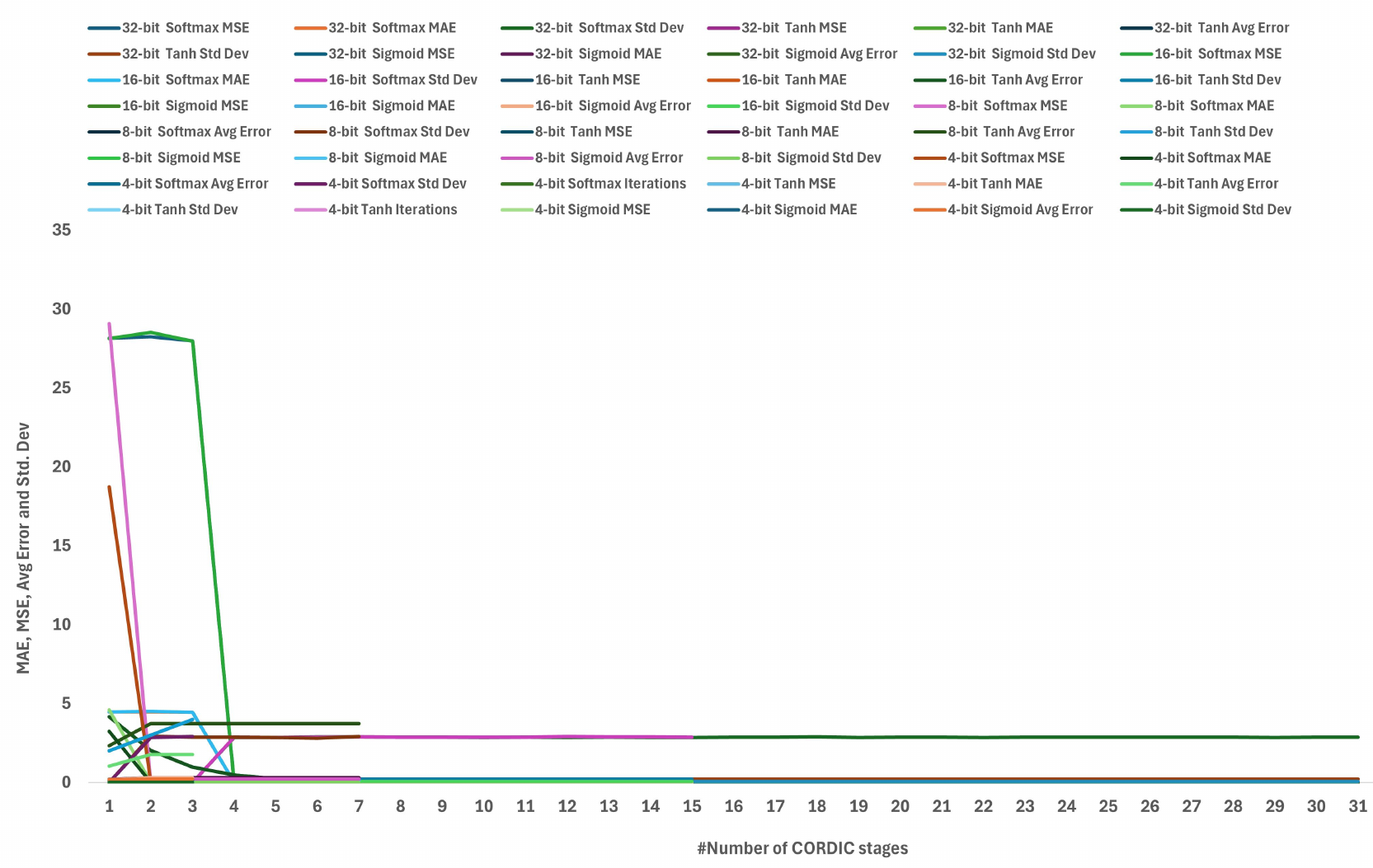}
        \caption{Pareto Evaluation for error metrics with proposed config-AF (softmax, sigmoid, tanh) with different LR and HV CORDIC stages for Flex-PE.}
    \label{Pareto-analysis}
    \vspace{-2mm}
\end{figure}

\subsection{Pareto Analysis for Identifying Optimal CORDIC stages}\label{sec:pareto}

The comprehensive Pareto analysis has determined the optimal CORDIC stages for various CORDIC-based operations across different fixed-point precisions: 4, 8, 16 and 32-bits.
The analysis provides valuable insights into the required stages for both pipelined and iterative PE implementations.
In pipelined mode, the redundant stages save hardware resources directly, while in iterative mode, the redundant clock cycles help reduce computational delay.
Our approach is similar to the previous approach, where the Pareto analysis is evaluated for RECON MAC~\cite{RECON}, sigmoid, tanh~\cite{GR-configAF-ISVLSI20} and Softmax~\cite{Softmax-Sumi}.
This helps balance area-delay trade-offs in fixed-point CORDIC designs based on available hardware resources and application performance.
The analysis revealed that 8-bit and 16-bit operations provide optimal performance for four HV stages (exponential) and five LR (division) and LV (MAC) stages, while for 32-bit operations, they increase to eight and nine/ten stages, respectively.
The Pareto analysis for 4-bit provides no benefits, thus utilising full 4-stage hardware.
However, the 32-bit pipeline hardware enables vertically time-multiplexed reconfigurability as almost half the stages are required for 8/16-bit operations, increasing throughput further by 2$\times$.
The point to be noted here is this is an additional enhanced throughput benefit apart from SIMD benefits.
The error-tolerant behaviour of AI applications can negate the minor impact of pareto analysis concerning mean errors introduced by CORDIC approximation.

\begin{figure}
    \centering
    \subfloat[]{\includegraphics[width=\columnwidth,height=30mm]{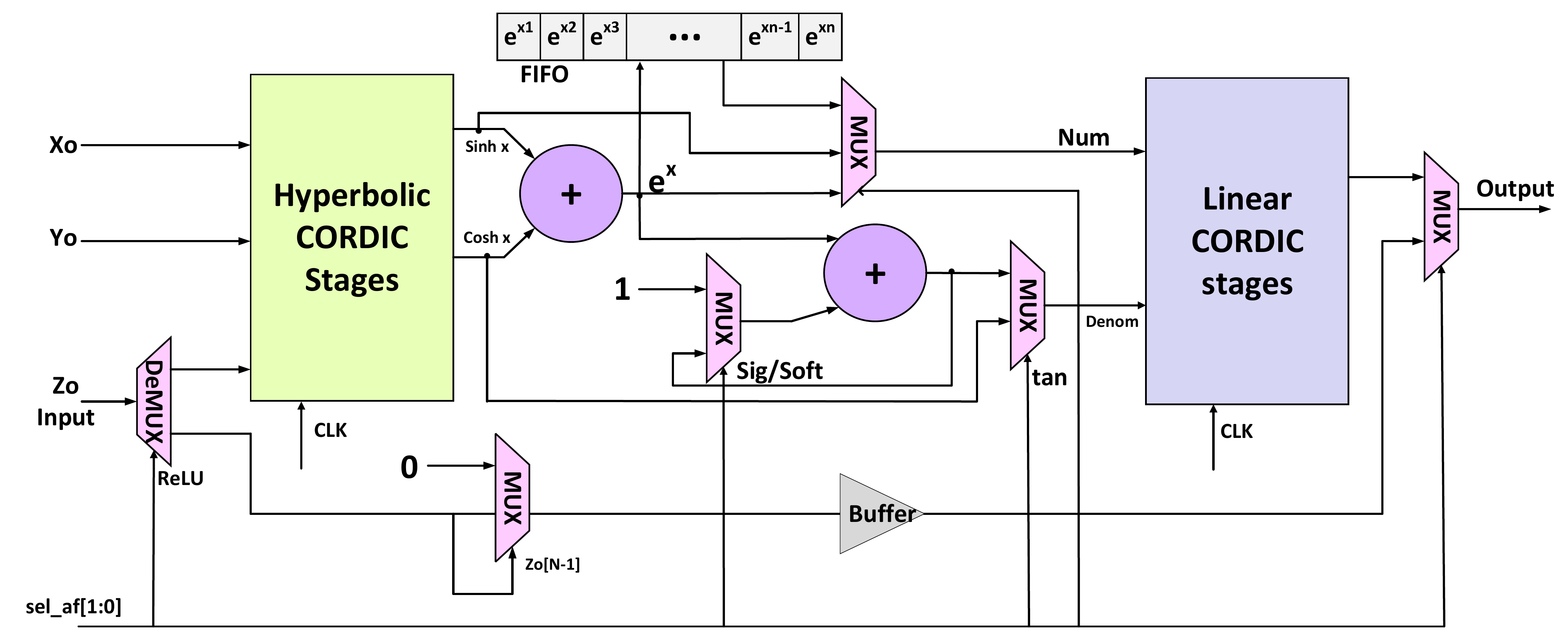}%
    \vspace{-3 mm}
    \label{SIMD-configAf}}
    \vspace{-2 mm}
    \hfill
    \subfloat[]{\includegraphics[width=\columnwidth,height=60mm]{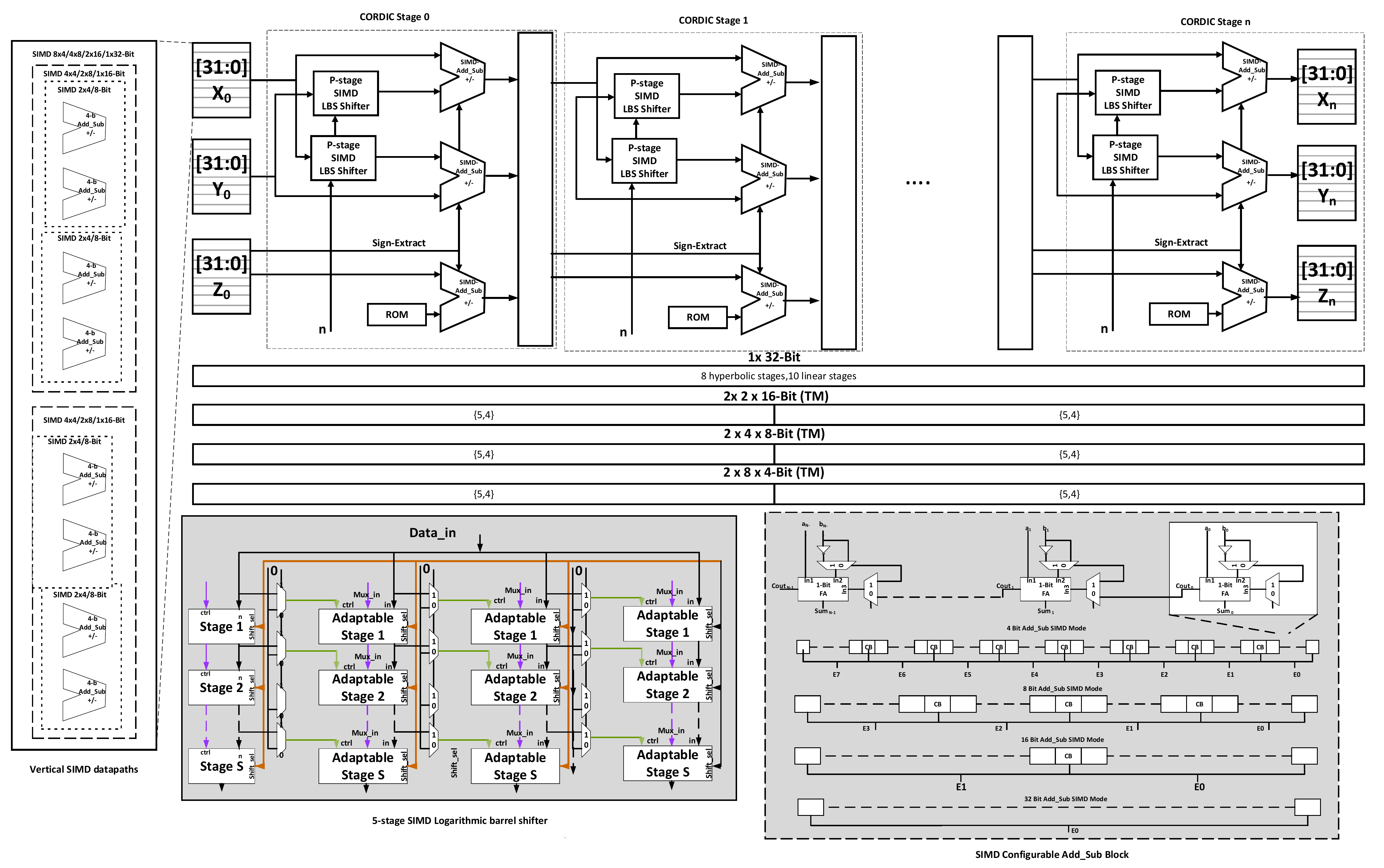}}%
    \label{detailed-PE}
    \caption{(a) Proposed SIMD FxP4/8/16/32 Configurable AF (Sigmoid, Tanh, ReLU, Softmax), (b) Detailed internal circuitry showcasing 5-stage SIMD Logarithmic barrel shifter and configurable Add\_Sub circuit design.}
    \label{flex-pe}
    \vspace{-5 mm}
\end{figure}

\section{Proposed SIMD re-configurable PE and Evaluation for multi-precision Systolic array}
The proposed work introduces novel SIMD-enabled CORDIC-based Flex-PE, which performs as SIMD MAC as well as runtime-configurable activation functions in 4/8/16/32-bit precision. The architecture leverages time-multiplexed pipelining techniques merged with precision-quantization precision to enhance throughput while dynamically maintaining area efficiency.
Conventional configurable AFs either support CORDIC-based Sigmoid/Tanh~\cite{GR-configAF-ISVLSI20, TC23-CORDIC-LSTM} or Taylor series and Newton Ralphson Division-based Sigmoid/Tanh/Swish activation functions~\cite{RECONFIG-MP-QuantAware-NAF, Softmax-taylor-DNN}.
However, these designs neither handle all the aforementioned activation functions nor support multi-precision SIMD 4/8/16/32 operations.
To tackle both issues, we propose a novel reconfigurable SIMD activation function with SIMD 4, 8, 16, 32-bit Add\_Sub block and 5-stage SIMD logarithmic barrel shifter with muxed logic to support time-multiplexing, as shown in Fig. \ref{flex-pe}.
This helps us achieve enhanced throughput of 16$\times$4-bit, 8$\times$8-bit, 4$\times$16-bit and 1$\times$32-bit AF operations. 

\begin{figure}[!t]
    \centering
    \includegraphics[width=\columnwidth,height=50mm]{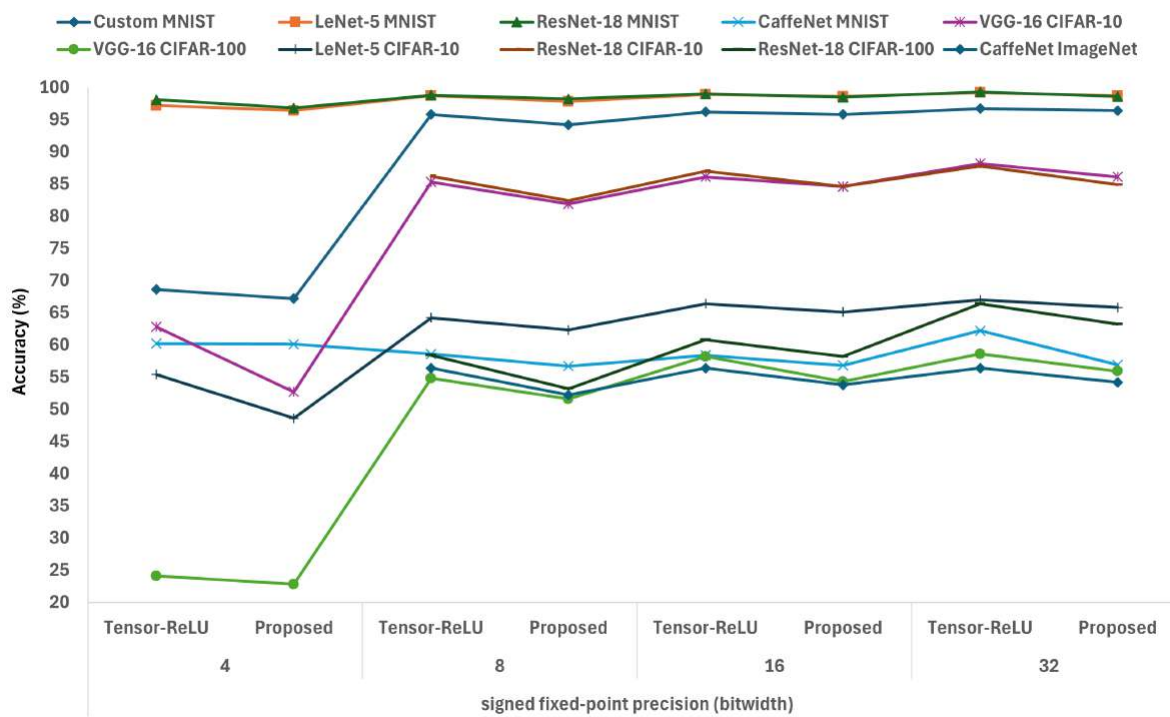}
        \caption{Evaluation of DNN accuracy showcasing effects of precision scalability on CORDIC-based SIMD processing engine (Flex-PE).}
    \label{Python-Accuracy}
    \vspace{-2mm}
\end{figure}

\subsection{Configurability}
The proposed SIMD FxP4/8/16/32 configurable AF is shown in Fig.4(a) with detailed data \& control signals. These signals help the control engine adaptively handle precision modes and/or different AFs. The architecture extracts enhanced performance and energy efficiency with the help of a dynamic control mechanism based on application performance demand and resources available. The mechanism also reduces overhead by eliminating separate datapaths like prior works and, by choosing on-the-fly AFs for diverse AI workloads. 
The Flex-PE uses SIMD CORDIC hyperbolic mode to calculate $\sinh$ and $\cosh$, as discussed in Section~\ref{sec:cordic-hyperbolic}. Further, with Muxed-logic, based on the Sel\_AF signal, the outputs are either directly forwarded for Tanh AF or added as per precision\_sel signal, and exponential computations can be performed. The selection between MAC and AF mode can be done with the help of the ctrl\_op signal. The ReLU function is implemented with simple mux-based reconfigurable logic from LBS shifters. 
To differentiate between Sigmoid and Softmax, muxed\_logic between MSB of Sel\_AF logic helps to propagate the previous stored exponential computations from FIFO or ``1". Furthermore, with SIMD division hardware, the outputs are calculated as soon as both operands are loaded.

\subsection{Run-time precision-variable SIMD AF}

In this work, we proposed the first fixed-point 4-bit Configurable Sigmoid/Tanh, beside ReLU for edge inference. It will play an integral role in accelerating workloads like RNN/LSTM and transformer inference. 
A novel SIMD-enabled 5-stage logarithmic barrel shifting (LBS) unit is implemented to handle simultaneous multi-precision FxP precision operations.
It also supports data parallelised rounds-to-even mode for Flex-PE.
Further, we developed a SIMD ripple carry chain-based addSub block for fixed-point arithmetic across multiple precisions, a critical block of CORDIC architectures.
Utilising FxP4 AF with muxed-logic with SIMD Add\_Sub block~\cite{Tapered-AddSub_ACM23} and novel 5-stage SIMD Logarithmic barrel shifter, we proposed a multi-precision FxP 4/8/16/32 configurable iterative Sigmoid/Tanh/ReLU and 8/16/32 configurable pipelined Sigmoid/Tanh/ReLU/Softmax with enhanced throughput of 16/8/4/1 by time-multiplexing the FxP32 hardware.
The proposed SIMD AF can also support 2$\times$FxP 12, 4$\times$FxP4 and 1$\times$FxP24 heterogenous operations.
The detailed internal circuitry is shown in Fig. 4 (b). 
The pipelined Flex-PE loads the inputs in two clock cycles without affecting bandwidth, and thus, outputs are produced alternatively in two clock cycles.
With SIMD AF, the proposed systolic accelerator can run parallel pipelines of workloads instead of conventional AF-bottlenecked AI hardware architectures~\cite{GR-Neuro, Intel-MxCore, Systolic-Trans-Fxp-TAI24}. 

\begin{table*}[]
    \caption{Comparative analysis of activation function units: Revealing FPGA resource utilization and performance metrics\\ for proposed FlexPE with pipelined Config-AF (Sigmoid/Tanh/Softmax/ReLU)}
    \label{Config-Soft-util}
    \resizebox{\textwidth}{!}{%
    \begin{tabular}{|c|cccccc|cccc|}
    \hline
    \textbf{AF} & \multicolumn{6}{c|}{\textbf{Softmax}} & \multicolumn{4}{c|}{\textbf{Proposed pipelined config-AF}} \\ \hline
    \textbf{Precision} & \textbf{FP32~\cite{MRao-ISQED24}} & \textbf{FP16~\cite{MRao-ISQED24}} & \textbf{BF16~\cite{MRao-ISQED24}} & \textbf{TF32~\cite{MRao-ISQED24}} & \textbf{FxP8~\cite{8-bSoftmax_APCCAS18}} & \textbf{FxP16~\cite{Zhu-16bSoftmax-TCASII20}} & \textbf{FxP8} & \textbf{FxP16} & \textbf{FxP32} & \textbf{SIMD FxP8/16/32} \\ \hline
    \multicolumn{11}{|c|}{\textbf{FPGA Utilization (VC707, 100 MHz)}}\\
    \hline
    \textbf{LUTs} & 3217 & 1137 & 1263 & 1259 & 1858 & 2564 & 256 & 427 & 681 & 897 \\ \hline
    \textbf{FFs} & - & - & - & - & 2086 & 2794 & 224 & 369 & 745 & 1231 \\ \hline
    \textbf{Delay(ns)} & 91.94 & 43.98 & 45.09 & 44.5 & 3.4 & 2.29 & 5.98 & 6.49 & 7.34 & 11.76 \\ \hline
    \textbf{Power (mW)} & 0.115 & 0.115 & 0.077 & 0.112 & - & - & 9.232 & 11.76 & 31.14 & 59.2 \\ \hline
    \textbf{PDP (pJ)} & 10.58 & 5.09 & 3.47 & 4.98 & - & - & 55.2 & 76.38 & 228 & 696 \\ \hline
    \multicolumn{11}{|c|}{\textbf{ASIC Utilization(28nm, 0.9 V)}}\\
    \hline
    \textbf{Area(um\textsuperscript{2})} & 41536 & 17289 & 11301 & 21927 & 15000 & 18392
    & 8570 & 20311 & 36153 & 49152 \\ \hline
    \textbf{Delay(ns)} & 5.95 & 3.97 & 3.3 & 4.52 & 2.23 & 0.31
    & 0.68 & 0.83 & 1.76 & 2.3 \\ \hline
    \textbf{Power (mW)} & 74.8 & 40.24 & 25.37 & 49.43 & 6.87 & 51.6
    & 1.5 & 2.43 & 3.37 & 5.1 \\ \hline
    \textbf{PDP (nJ)} & 0.44 & 0.16 & 0.084 & 0.23 & 15.32 & 16
    & 1.02 & 2 & 5.93 & 11.96 \\ \hline
    \end{tabular}}
    \vspace{-3mm}
\end{table*}

\begin{table*}[!t]
    \caption{Comparative analysis of activation function units: Revealing FPGA resource utilization and performance metrics \\ for proposed FlexPE with iterative Config-AF(Sigmoid/Tanh/ReLU)}
    \label{fpga-config1}
    \resizebox{\textwidth}{!}{%
    \begin{tabular}{|c|cccc|cccc|ccccc|}
    \hline
    \textbf{AF} & \multicolumn{4}{c|}{\textbf{Tanh}} & \multicolumn{4}{c|}{\textbf{Sigmoid}} & \multicolumn{5}{c|}{\textbf{Proposed Iterative config-AF}}\\ \hline
    \textbf{Precision} & \textbf{FP32~\cite{MRao-ISQED24}} & \textbf{FP16~\cite{MRao-ISQED24}} & \textbf{BF16~\cite{MRao-ISQED24}} & \textbf{TF32~\cite{MRao-ISQED24}} & \textbf{FP32~\cite{MRao-ISQED24}} & \textbf{FP16~\cite{MRao-ISQED24}} & \textbf{BF16~\cite{MRao-ISQED24}} & \textbf{TF32~\cite{MRao-ISQED24}} & \textbf{FxP4} & \textbf{FxP8} & \textbf{FxP16} & \textbf{FxP32} & \textbf{SIMD FxP4/8/16/32} \\ \hline
    \multicolumn{14}{|c|}{\textbf{FPGA Utilization (VC707, 100 MHz)}}\\
    \hline
    \textbf{LUTs} & 4298 & 1530 & 1513 & 1990 & 5101 & 1853 & 1856 & 2436 & 45 & 84 & 140 & 257 & 405 \\ \hline
    \textbf{FFs} & - & - & - & - & - & - & - & - & 37 & 72 & 126 & 221 & 116 \\ \hline
    \textbf{Delay(ns)} & 56.6 & 34 & 37.8 & 42.2 & 109 & 60.6 & 44.5 & 61.7 & 0.91 & 1.57 & 2.21 & 2.77 & 3.8 \\ \hline
    \textbf{Power (Logic + Signal) } & 0.13 & 0.124 & 0.082 & 0.118 & 0.121 & 0.118 & 0.083 & 0.116 & 2 & 10.26 & 14.95 & 22 & 35 \\ \hline
    \textbf{PDP (pJ)} & 7.35 & 4.22 & 3.1 & 4.98 & 13.25 & 7.16 & 3.69 & 7.15 & 1.82 & 16.12 & 33.05 & 60.94 & 133 \\ \hline
    \multicolumn{14}{|c|}{\textbf{ASIC Utilization (28 nm, 0.9 V)}}\\
    \hline
    \textbf{Area (um\textsuperscript{2})} & 5060 & 1180 & 843 & 1728 & 2234 & 1855 & 1180 & 2234 
    & 78 & 132 & 313 & 751 & 987 \\ \hline
    \textbf{Power (mW)} & 8.75 & 3.06 & 2.08 & 4.2 & 10.06 & 4.81 & 2.45 & 5.42 
    & 0.13 & 0.19 & 0.31 & 0.42 & 0.73 \\ \hline
    \textbf{Delay(ns)} & 3.92 & 3.31 & 3.38 & 3.47 & 7.58 & 4.37 & 3.26 & 4.34 
    & 2.86 & 3.98 & 4.32 & 5.82 & 7.34 \\ \hline
    \textbf{PDP (pJ)} & 34.3 & 10.2 & 7.03 & 14.6 & 76.35 & 20.97 & 7.95 & 23.76 
    & 0.37 & 0.76 & 1.34 & 2.43 & 5.35 \\ \hline
    
    \end{tabular}}
\end{table*}

\subsection{Flex-PE for wide-range AI workloads}

The proposed Flex-PE hardware exploits the Quantized SIMD operations in workloads over diverse precision ranges of 4, 8, 16, 32-bits to support a wide range of workloads from edge inference and RNN/LSTM to cloud HPC applications, such as DNN training and Transformers, to efficiently enhance the performance.
The same hardware used in the computation of AFs can be rewired to support MAC computations, similar to RECON~\cite{RECON, GR-ACM_TRETS23}, and enables a more area-efficient AI hardware solution for DNN/Transformers~\cite{Systolic-array-CNN, Lai-TVLSI19, Systolic-Trans-Fxp-TAI24}.
In our knowledge, such versatility within a single Flex-PE is the first of its kind in SoTA works.
The processing element can be very efficient as a fundamental block for systolic array architectures, as shown in Fig. 1 (b), enhancing the throughput with multi-precision operations. The approach helps unprecedented versatility and flexibility to adapt to dynamically changing computational and memory constraints. The processing element is enhanced with the support for FxP 4/32 precision in addition to prior work, significantly broadening the range of application performance. It delivers superior compute density and energy efficiency for resource-constrained execution while scaling easily to higher precision if required for HPC computations.

\section{Implementation Methodology and Performance Analysis}

The experimental setup for evaluating the proposed methodology includes software-based workload analysis and hardware-based architectural emulations to ensure the codesign for the accelerator, which has been briefed here. 

We conducted a detailed Pareto analysis for Flex-PE, as discussed in Section~\ref{sec:pareto}, to determine the optimal Pareto stages required in the computations of various operations, such as MAC, Sigmoid AF, Tanh AF, and Softmax AF, using CORDIC design methodology.
The analysis provides detailed trade-offs between the number of Pareto stages required, error metrics and computation delay of iterative config-AF or Area efficiency of pipelined config-AF.
This also ensures the enhancement in throughput within optimal area and energy consumption.
The Pareto points can be extracted from error deviation (MAE and MSE) and optimal accuracy when varied with the number of CORDIC stages.
The Monte Carlo error simulations for uniformly distributed $2^{(N/2) + 1}$ times for 4, 8, 16 and 32-bit precision on random input pattern.
The simulations were compared with true outputs from Python Numpy, and mean square error (MSE) and mean average error (MAE) metrics would give justifiable approximation trade-offs.

Our analysis revealed that 8-bit and 16-bit operations provide optimal performance for four HV stages (exponential) and five LR (division) and LV (MAC) stages. In contrast, 32-bit operations increase to eight HV and nine/ten LR/LV stages, respectively.
The Pareto analysis for 4-bit provides no benefits, thus utilizing full 4-stage hardware. The detailed analysis is showcased in Fig. \ref{Pareto-analysis}. 
The number of CORDIC stages can be increased up to a maximum of $N$ for $N$-bit computations to minimize the error by sacrificing the computing and throughput benefits of hardware resources.

\begin{table*}[!t]
\caption{COMPARATIVE FPGA IMPLEMENTATION ANALYSIS OF ACTIVATION FUNCTION UNITS with state-of-the-art works}
\label{SOTA-FPGA}
\resizebox{\textwidth}{!}{%
\begin{tabular}{|c|cc|c|c|c|c|cc|}
\hline
\textbf{Parameter} & \multicolumn{2}{c|}{\textbf{TCAD'19\cite{TCAD19-AF}}} & \textbf{TCAS-II'23\cite{TCASII-23_ReAFM}} & \textbf{TVLSI'23\cite{Softmax-taylor-DNN}} & \textbf{TC'23\cite{TC23-CORDIC-LSTM}} & \textbf{TAI'24\cite{TAI24-CORDIC-RNN}} & \multicolumn{2}{c|}{\textbf{Proposed}} \\ \cline{2-9} 
 & \multicolumn{1}{c|}{\textbf{Tanh-8b}} & \textbf{SELU-8} & \textbf{Sigmoid/Swish/Tanh-12b} & \textbf{Softmax-16b} & \textbf{Tanh/Sigmoid-16b} & \textbf{Tanh/Sigmoid} & \multicolumn{1}{l|}{\textbf{SSTp-8/16/32}} & \multicolumn{1}{l|}{\textbf{SSTi-4/8/16/32}} \\ \hline
\textbf{FPGA} & \multicolumn{1}{c|}{VC707} & VC707 & XC7VX-690T & Zynq-7 & Pynq-z1 & Zynq7000 & \multicolumn{1}{c|}{VC707} & VC707 \\ \hline
\textbf{LUT} & \multicolumn{1}{c|}{-} & - & 367 & 1215 & 36286 & 2395 & \multicolumn{1}{c|}{897} & 405 \\ \hline
\textbf{FF} & \multicolumn{1}{c|}{-} & - & 298 & 1012 & 24042 & 1503 & \multicolumn{1}{c|}{1231} & 116 \\ \hline
\textbf{Power (mW)} & \multicolumn{1}{c|}{84} & 104 & - & 165 & 125 & 0.681 & \multicolumn{1}{c|}{59.2} & 3.8 \\ \hline
\textbf{Delay (us)} & \multicolumn{1}{c|}{-} & - & 0.35 & 3.32 & 21 & 0.18 & \multicolumn{1}{c|}{11.76} & 35 \\ \hline
\end{tabular}}
\end{table*}

\begin{table*}[!t]
\caption{COMPARATIVE ASIC IMPLEMENTATION ANALYSIS OF ACTIVATION FUNCTION UNITS with state-of-the-art works}
\label{SoTA-ASIC}

\resizebox{\textwidth}{!}{%
\begin{tabular}{|c|cc|cc|c|c|c|c|cc|}
\hline
\textbf{Parameter} & \multicolumn{2}{c|}{\textbf{TCAD'19\cite{TCAD19-AF}}} & \multicolumn{2}{c|}{\textbf{OJSCAS'21\cite{RECON}}} & \textbf{TC'23\cite{TC23-CORDIC-LSTM}} & \textbf{TCAS-I'23\cite{TCASI23-Softmax}} & \textbf{TVLSI'23\cite{Softmax-taylor-DNN}} & \textbf{TCS-VT'24\cite{TCSVT24_SoftAct}} & \multicolumn{2}{c|}{\textbf{Proposed}} \\ \cline{2-11} 
 & \multicolumn{1}{c|}{\textbf{Tanh-8b}} & \textbf{SELU-8} & \multicolumn{1}{c|}{\textbf{Sigmoid/Tanh-8b}} & \textbf{Tanh/Sigmoid-16b} & \textbf{Tanh/Sigmoid-16b} & \textbf{Softmax-32} & \textbf{Softmax-16b} & \textbf{Softmax-16b} & \multicolumn{1}{l|}{\textbf{SST-8/16/32}} & \multicolumn{1}{l|}{\textbf{SST-4/8/16/32}} \\ \hline
\textbf{ASIC Tech (nm)} & \multicolumn{1}{c|}{28} & 28 & \multicolumn{1}{c|}{45} & 45 & 45 & 28 & 28 & 28 & \multicolumn{1}{c|}{28} & 28 \\ \hline
\textbf{Area (um2)} & \multicolumn{1}{c|}{97.65} & 138 & \multicolumn{1}{c|}{794} & 24608 & 870,523 & 98787 & 3819 & 0.0028 & \multicolumn{1}{c|}{49152} & 987 \\ \hline
\textbf{Power (mW)} & \multicolumn{1}{c|}{-} & - & \multicolumn{1}{c|}{526} & 1033 & 151 & 24.72 & 1.58 & 3.46 & \multicolumn{1}{c|}{5.1} & 0.73 \\ \hline
\textbf{Delay (ns)} & \multicolumn{1}{c|}{0.195} & 0.22 & \multicolumn{1}{c|}{3.73} & 4.76 & - & 26 & 1.6 & - & \multicolumn{1}{c|}{2.3} & 7.34 \\ \hline
\textbf{Max Op freq (GHz)} & \multicolumn{1}{c|}{5.14} & 4.52 & \multicolumn{1}{c|}{5} & - & - & 1 & - & 1.85 & \multicolumn{1}{c|}{5} & 5 \\ \hline
\end{tabular}}
\end{table*}

We designed the proposed CORDIC methodology-based model using Python 3.0 and QKeras 2.4 library using the Google Colab platform for 4, 8, 16 and 32-bit FxP precision, precisely similar to hardware arithmetic design.
Unlike prior works~\cite{GR-ACM_TRETS23, GR-ISQED24}, our model evaluated the accuracy using ResNet-18, VGG-16 on the CIFAR-100 dataset, with purely CORDIC-based MAC, Sigmoid/Tanh and Softmax activation functions (SST) in classification layer.
We also evaluated smaller custom CNN, LeNet-5 models for edge inference and proved that the accuracy loss is negligible (\(<2\%\)) compared with standard TensorFlow arithmetic, as depicted in Fig. \ref{Python-Accuracy}.
This ensures the proposed CORDIC-based approach maintains accuracy within 98\% Quality of Results (QoR).

Based on the detailed Pareto points discussed, the error analyses (MAE, MSE) and ASIC analysis to identify the optimal number of pipelined or iterative stages to achieve the best trade-off for optimum accuracy with hardware performance. We have implemented the FSM in the control engine for iterative mode and the number of CORDIC stages in pipelined mode. 
We further described the proposed SIMD iterative config-AF, SIMD pipelined config-AF, Flex-PE and multi-precision systolic array with Verilog-HDL language for different fixed-point precisions. Furthermore, we simulated the results with the Questa-Sim simulator and cross-validated them with the python-emulation framework.

We performed the FPGA synthesis and implementation with the AMD Vivado Design Suite, and post-implementation resources were reported. Table \ref{Config-Soft-util} discusses the comparative analysis for proposed Flex-PE with pipelined config-AF, with significant improvement in LUTs and FFs. The resource utilization for Flex-PE with iterative config-AF is reported in Table \ref{fpga-config1}. The pipelined design achieves high throughput due to the presence of feedback registers. The iterative design trade-off reduction in area for latency-induced time-multiplexed computation. The comparison between iterative and pipeline CORDIC stages shows a 5$\times$ area reduction and delay in the inverse case. We have also compared it with floating-point precision hardware to highlight the importance of hardware savings without compromising significant accuracy. The performance comparison between different hardware architectures, such as layer-reused\cite{GR-ACM_TRETS23}, flexible NullHop\cite{NullHop-FlexAccl-TNNLS19}, sparse systolic accelerator\cite{RAMAN-IoTJ24, Zhu-SparseCNN-TVLSI20}, etc. is discussed with SoTA comparison in the Table \ref{HW-arch}. 
We synthesized this design using CMOS 28nm technology with Synopsys Design Compiler, and post-synthesis parameters for the proposed SIMD iterative config-AF, SIMD pipelined config-AF, are described in Table \ref{Config-Soft-util} and Table \ref{fpga-config1}. We also resynthesized SoTA designs using the same technology node to maintain similar parameters for a fair comparison.
The FPGA and ASIC implementations of the proposed Iterative config-AF and Pipelined config-AF showcase significant improvements over conventional floating-point AFs and SoTA works, with the additional advantage of computing MAC computations within the same Flex-PE.

\vspace{-3mm}

\subsection{Evaluation for SIMD Systolic array}

Emerging AI applications like AR/VR, autonomous navigation systems and AI-driven applications depend heavily on generalised systolic array hardware accelerators as crucial fundamental hardware blocks.
Efficient unified programmable hardware~\cite{Intel-MxCore} is required to support various applications and diverse matrix computations in resource-constrained environments.
We build a RISC-V-enabled SIMD systolic architecture, as depicted in Fig.
\ref{Edge-AI-SoC}, to evaluate the impact of the proposed multi-precision PE with configurable AF. The systolic array accelerator can be easily interfaced with AXI and DMA of Cheshire\cite{Cheshire} and validated with the help of p-type SIMD API calls. Cheshire is a lightweight, linux-capable RISC-V host platform for accelerator plug-in in AMD Vivado Design Suite. The synchronization is handled with the help of status registers, control signals and a custom systolic array control engine.
The architecture is highly scalable, and array size can be varied in run-time; thus, for simplicity of our implementation, we validated the 8$\times$8 array. 
The systolic accelerator enables all general matrix multiplication applications combined with configurable AFs suitable for all AI applications, such as DNN/RNN, Transformers and reduces the data movement between SIMD programmable Matrix Block to SISD multi-threaded programmable cores.
Furthermore, with the support of diverse activation functions, such as ReLU, Sigmoid, Tanh, Softmax, and MAC computations in 4, 8, 16 and 32-bit-precisions, the proposed accelerator can benefit from many SoTA hardware-software co-design techniques \cite{Systolic-array-CNN, Systolic-Trans-Fxp-TAI24} to elevate the performance for DNN/transformer applications.

\begin{table*}[!t]
    \caption{Hardware Implementation Report with Proposed Systolic Array Architecture and State-of-the-Arts DNN Designs}
    \label{HW-arch}
    \resizebox{\textwidth}{!}{%
    \begin{tabular}{cccccccccc}
    \hline
    \textbf{} & \textbf{Platform} & \textbf{Model} & \textbf{Precision} & \textbf{LUTs (Thousands)} & \textbf{Registers (Thousands)} & \textbf{DSPs} & \textbf{Op.
    Freq (MHz)} & \textbf{Energy efficiency (GOPS/W)} & \textbf{Power (Watts)} \\ \hline
    \textbf{TRETS'23~\cite{GR-ACM_TRETS23}} & VC707 & Custom & 8/16 & 115 & 115 & 32 & 100 & 4.5 & 2 \\ 
    \textbf{Neuro'22~\cite{GR-Neuro}} & VC707 & Custom & 8 & 210 & 310 & 57 & 200 & 8 & 4 \\ 
    \textbf{TNNLS'19~\cite{NullHop-FlexAccl-TNNLS19}} & Zynq7 & VGG16 & 16 & 229 & 107 & 128 & 60 & 27.5 & 1.1 \\
    \textbf{IoTJ'24~\cite{RAMAN-IoTJ24}} & Ti60 & MobileNetV1 & 2/4/8 & 37.2 & 8.5 & 61 & 75 & 98.5 & 0.14 \\ 
    \textbf{TVLSI'20~\cite{Zhu-SparseCNN-TVLSI20}} & ZCU102 & ResNet-50 & 16 & 390 & 278 & 1352 & 200 & N/A & 15.5 \\ 
    \textbf{TCAS-I'21~\cite{Xie-FlexAccl-TCASI21}} & Arria10 & MobileNetV2 & 8 & 102.5 & N/A & 512 & 170 & 18.7 & 4.6 \\ 
    
    \textbf{TCAS-I'22~\cite{Lu-SparseCNN-TCASI22}} & ZC706 & 1-D CNN & 16 & 3.24 & N/A & 48 & 200 & 45 & 0.5 \\     
    \textbf{ESL'24~\cite{Yin-StrucSparse-ESL24}} & ZCU102 & MobileNetV2 & 16 & 194.5 & 95.7 & 884 & 190 & N/A & 13.3 \\    
    \textbf{TCAS-I'24~\cite{Wu-Sp-systolic-TCASI24}} & ZU3EG & ResNet-50 & 8 & 40.8 & 45.25 & 258 & 150 & 45 & 1.4 \\ 
    \textbf{Proposed } & VC707 & Custom & 4/8/16/32 & 38.72 & 7.4 & 73 & 466 & 8.42 & 2.24 \\ \hline
    \end{tabular}%
    }
\end{table*}

The detailed comparison with SoTA architectures can be analysed from Table \ref{HW-arch}.
Many SoTA architectures elaborate on the in-depth analysis of multi-precision MAC computations~\cite{Intel-MxCore, Multi-mode-MAC_JSSC23}, sparse matrix computations~\cite{Lai-DAC, slimmerCNN, RAMAN-IoTJ24}, with and/or Tiling-Sequencing ~\cite{Intel-MxCore, Systolic-PE_TCASI24, Lai-TVLSI19}.
Our approach enhances it by reducing on-chip data transfer involved in AF computations with enhanced throughput and power optimization at both edge and cloud nodes.
The proposed architecture further enhances the throughput by combining benefits from run-time quantization and systolic computation without affecting the memory bandwidth requirement.
It is much more efficient for parallelizable workloads and reduces execution latency with sparsity~\cite{Lai-DAC}.
The software codesign benefits are extracted with custom SIMD-enabled API calls~\cite{Systolic-array-CNN}, configuration registers, and a custom compilation framework.
AlexNet and VGG16 achieve 10$\times$ and 62$\times$ reductions in DMA reads for input fmaps and 214$\times$ and 371$\times$ for weight filters, significantly reducing power consumption and latency with our SIMD-enabled API.
The pipelined dataflow minimizes memory access, especially for convolution and multi-headed attention operations with data reuse.

\begin{figure}[!t]
    \centering
    \includegraphics[width=\columnwidth,height=50mm]{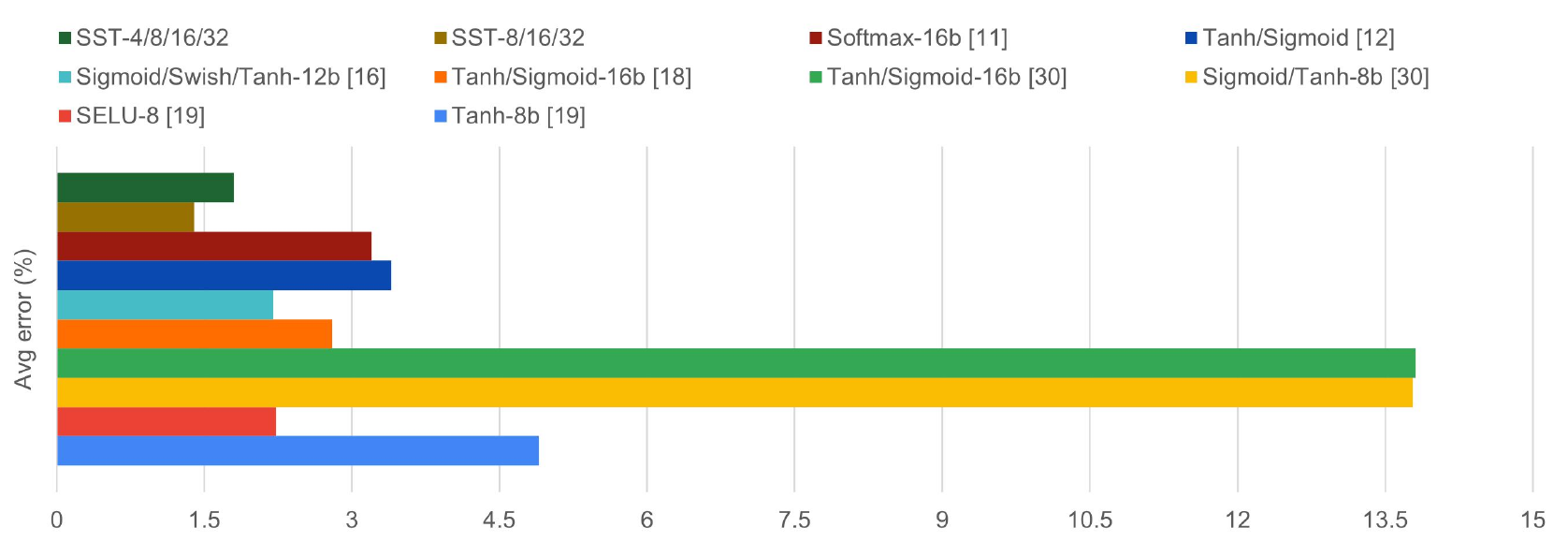}
        \caption{Comprehensive Evaluation of Mean Error metrics of SSTp (proposed pipelined config-AF) and SSTi (proposed iterative config-AF) with SoTA works.}
    \label{mean-error}
    \vspace{-2mm}
\end{figure}

The detailed FPGA implementation results are compared in Table \ref{HW-arch} and showcase the improvement in the operating frequency of the proposed design while reducing the hardware resource due to reconfigurable MAC and AF. The proposed SIMD SA enables efficient computations from a wide range of precision and incorporates a run-time switch between different AFs, enhancing the SIMD Systolic array hardware to support diverse workloads. It supports DNN workloads with MAC and ReLU operations. For sequential tasks, such as RNNs and LSTM models, it iteratively handles adaptive workloads with sigmoid and tanh activation functions. Further, it enhances the performance of transformer applications with a huge number of parallel reconfigurable softmax hardware. The comprehensive support of diverse precision-based throughput emphasises versatility for 4-bit edge inference to 32-bit HPC computations.

\subsection{Application scenarios \& Future work}

The precision modes FxP4 and FxP8 are better suited for faster inference tasks, offering speedup with enhanced throughput for AI workloads. 
Higher precision modes cater to the needs of large-scale cloud training, perfectly managing accuracy and computational demands.
The 4-bit computations achieve 16$\times$ throughput with a slight loss in accuracy \(<2\%\), which is acceptable for major AI applications. 
The iterative design optimizes the area for edge Applications, whereas pipelined designs offer enhanced throughput based on workload requirements. 
The system can provide quick inference within minimal resources for autonomous systems with 4-bit precision; however, adjusting critical layers with higher precision avoids minimum performance deterioration. The mean error for prior works with this work has been compared in Fig \ref{mean-error}. 
The FxP32 precision shall be explored more for training purposes on edge devices for diverse real-world applications. The work could be considered a strong foundation for future AI accelerator architectures, with huge potential for high-performance, energy-efficient artificial general intelligence hardware.
The scope of this work is limited to the above-mentioned AFs and can be easily extended to activation functions, such as Swish and GELU, with the same CORDIC hardware.

\section{Conclusion}

The proposed FLEX-PE introduces a first-of-its-kind, runtime reconfigurable activation function with SIMD multi-precision processing and achieves throughput enhancement up to 16× for FxP4, 8× for FxP8, and 4× for FxP16 with 100\% time-multiplexed pipeline hardware. It functions as both SIMD MAC and diverse AF, enabling scalable neuron design for addressing the computational demands with the reduction in underutilized dark silicon. Flex-PE provides superior performance for state-of-the-art AI workloads, ranging from edge inference to HPC applications, with runtime switches between key activation functions (Sigmoid, Tanh, ReLU, and Softmax). The proposed CORDIC-based systolic array solution reduces DMA accesses for input feature maps and weight filters in VGG-16 up to 62× and 371× with SIMD API, with remarkable energy efficiency of 8.42 GOPS/W with less than 2\% accuracy loss. Compared to state-of-the-art solutions, Flex-PE provides programmable hardware supported for diverse AI workloads, including DNNs, RNNs, LSTMs, and Transformers.

The proposed design adapts to performance, power, and varying requirements in edge and cloud deployments. The area-efficient iterative multi-precision mode supports 4-bit computations for deep learning inference, while SIMD pipelined mode supports up to 32-bit computations for HPC requirements. The work establishes a strong foundation for developing future high-performance, energy-efficient artificial general intelligence hardware.

\newpage
\section{Biography Section}

\begin{IEEEbiography}[{\includegraphics[width=1in,height=1.25in, clip,keepaspectratio]{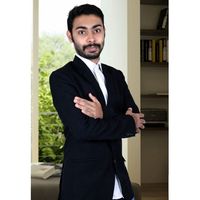}}]{Mukul Lokhande} (Graduate Student Member, IEEE) received his B.Tech. Degree in Electronics and Telecommunication Engineering from Shri Guru Gobind Singh Ji Institute of Engineering and Technology, Nanded, India, in 2020.

He is doctoral research scholar at the Department of Electrical Engineering, Indian Institute of Technology Indore, India. His research focuses on innovative hardware-software co-design techniques to improve the efficiency and performance of AI accelerators. Specifically, his work explores approximation methods, Quantization-enabled SIMD hardware, and CORDIC-based computation to optimize arithmetic-intensive tasks to bridge the gap between computational accuracy and resource efficiency, targeting AI applications.
\end{IEEEbiography}

\begin{IEEEbiography}[{\includegraphics[width=1in,height=1.25in, clip,keepaspectratio]{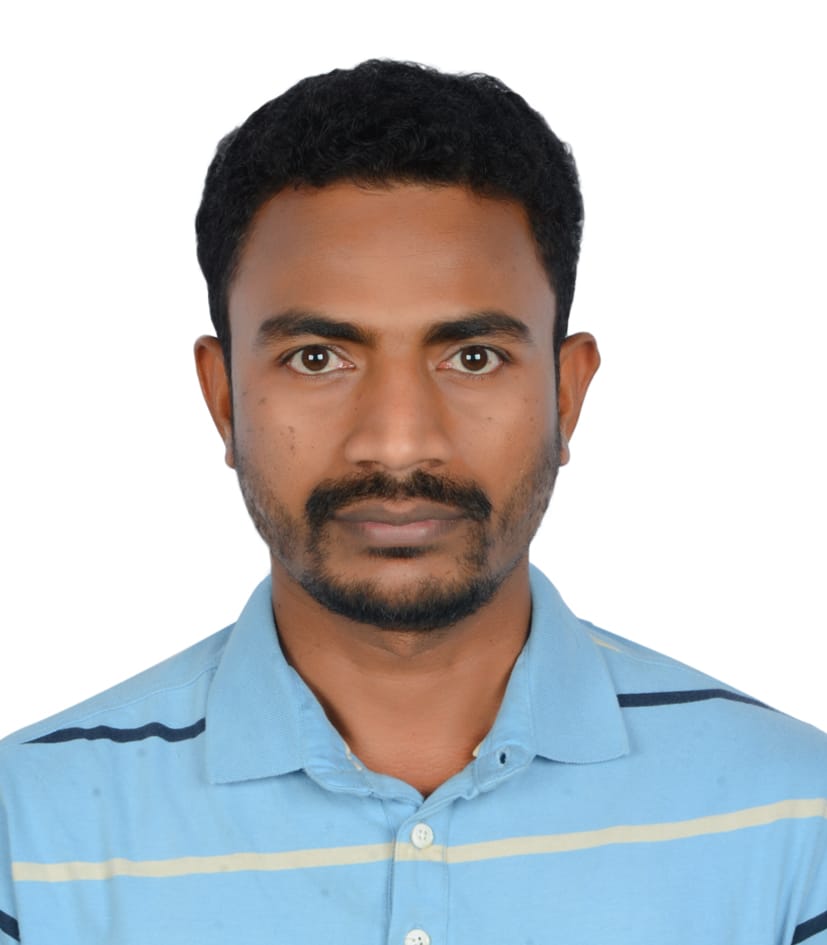}}]{Gopal Raut} (Member, IEEE) received the master’s degree in VLSI design from the G. H. Raisoni College of Engineering, Nagpur, India, in 2015, and the Ph.D. degree from the Indian Institute of Technology Indore, India, in 2022.

His research interests include compute-efficient design and hardware implementation of DNN accelerators for IoT and edge-AI applications.
\end{IEEEbiography}

\begin{IEEEbiography}[{\includegraphics[width=1in,height=1.25in, clip,keepaspectratio]{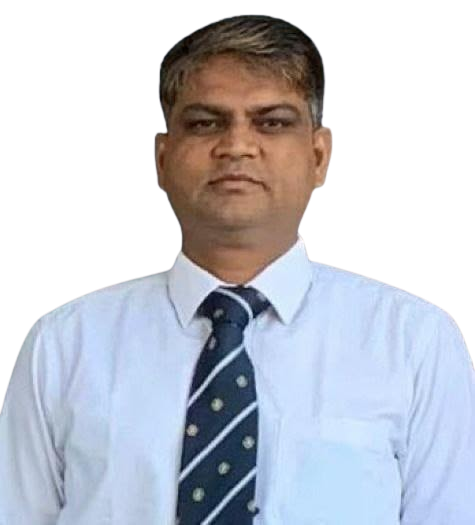}}]{Santosh Kumar Vishvakarma} (Senior Member, IEEE) received the Ph.D.Degree from the Indian Institute of Technology Roorkee, India, in 2010.
From 2009 to 2010, he was with the University Graduate Center, Kjeller, Norway, as a Postdoctoral Fellow under a European Union Project.
He is a Professor with the Department of Electrical Engineering, Indian Institute of Technology Indore, India, leading the Nanoscale Devices, VLSI Circuit and System Design Laboratory.
His research interests include nanoscale devices, reliable SRAM memory designs, and reconfigurable circuit design for edge AI applications.
\end{IEEEbiography}

\end{document}